\documentclass[12pt]{iopart}

\usepackage[numbers, sort&compress]{natbib} 
\usepackage{hyperref}

\hypersetup{
citecolor=blue,
linkcolor=blue,
colorlinks=true,
}
\usepackage{graphicx}

\begin{document}

\title[]{Optical tweezers assisted coupling of nematic droplets to gold nanoparticle cluster: effect on whispering gallery modes}

\author{Sumant Pandey, G V Pavan Kumar}

\address{Department of Physics, Indian Institute of Science Education and Research (IISER) Pune, Pune, 411008, India}

\ead{pavan@iiserpune.ac.in}
\vspace{10pt}
\begin{indented}
\item[]
\end{indented}

\begin{abstract} 
Dye-doped liquid crystal (LC) microdroplets exhibit tunable optical resonances modulated by size, shape, temperature, and external perturbations. When a dye-doped nematic microdroplet is coupled to a gold nanoparticle cluster, near-field interactions enhance local electric fields, boosting fluorescence emission. Optical tweezers serve as a tool for the parking of dye-doped nematic microdroplets on gold nanoparticle clusters, enabling the dynamic coupling and excitation of whispering-gallery modes (WGMs). This configuration resulted in amplified WGMs, with a clearly detectable shift in the spectral position. Resonance mode red shifts confirmed efficient photonic–plasmonic coupling, with up to 7 nm tunability achieved without significant degradation of the Q-factor. The magnitude of tunability depends on the size of the gold nanoparticle cluster. Also, the WGM emission spectrum of the nematic microdroplet can be reversibly tuned by decoupling from the gold nanoparticle cluster.
\end{abstract}

%

%
%
%

\section{Introduction}
Liquid crystals (LCs)\cite{de1993physics} exhibit self-assembly across molecular to macroscopic scales, characterised by anisotropic optical responses while preserving fluid-like mobility.\cite{bisoyi2021liquid, tschierske2013development, van2017liquid} Liquid crystal spherical droplets exhibit adaptive and stimuli-responsive behaviour along with photonic band gap characteristics, making them dynamic soft matter systems with photonic applications.\cite{miller2014design, broer2012functional, kato2018functional, humar20103d, muvsevivc2016liquid} In nematic droplets, the alignment direction that is sufficiently anchored at the boundary will be transmitted into the bulk in the form of an orientation field deformation, causing the generation of topological defects or dislocations.\cite{chen2021liquid} These defects are highly sensitive to changes in size and surfactant concentration\cite{shechter2020direct}, and their interfacial charge modulation has even been exploited for ultrasensitive protein detection \cite{maheshwari2024charge}. Nematic droplets are highly responsive interfacial systems that can be functionalized through diverse molecular interactions, exhibiting polymer–surfactant–driven structural modulation \cite{naveenkumar2021polymer} and enabling spatial control over protein organisation via acoustic wave trapping \cite{naveenkumar2023patterning}. 

These nematic microdroplets in confined spherical geometries act as optical microresonators supporting high-Q resonant modes.\cite{humar2009electrically} Whispering-gallery-mode (WGM) resonators confine light via total internal reflection, allowing it to circulate inside a spherical cavity and return to its original position in phase after each round trip.\cite{vahala2003optical} This circulation leads to significantly enhanced photon lifetimes\cite{qian1986lasing} and makes WGMs highly sensitive to even minute perturbations in the surrounding medium.\cite{datsyuk2001optics, giorgini2017fundamental} Dye-doped nematic microdroplets are particularly attractive for their tunable properties\cite{humar2009electrically, lee2015measuring, sofi2017electrical, kumar2015detection, sofi2019stability, venkitesh2024two, jonavs2017thermal}, which arise due to their elastic nature and are prominent candidates in chemical sensors\cite{humar2011surfactant, duan2020real}, microlasers \cite{humar2015intracellular, kiraz2007lasing, mur2017magnetic, sofi2019electrically, sofi2019whispering}. Tunability is one of these resonators' most appealing characteristics; it allows the resonance modes to be adjusted by varying the applied electric and magnetic fields, temperature, size and other perturbations.

Plasmonic nanostructures strongly enhance light–matter interactions via sub-wavelength modes\cite{kauranen2012nonlinear, tame2013quantum}, enabling effects such as surface-enhanced Raman scattering (SERS)\cite{ le2008principles, langer2019present, kumar2012plasmonic, lim2010nanogap, park2018quantitative} and surface-enhanced fluorescence (SEF)\cite{lakowicz2006principles, geddes2002metal, anger2006enhancement, sultangaziyev2020applications, fort2007surface, lakowicz2004advances}, but suffer from significant dissipative losses\cite{ahn2013demonstration, gramotnev2010plasmonics}. In contrast, WGM microresonators offer low optical losses and high Q-factors but exhibit relatively larger mode volumes, resulting in weaker local field enhancements. Integrating these complementary systems into hybrid structures combines the strengths of both the strong field localisation of plasmonics with the long photon lifetimes of WGMs, thereby enabling advanced applications in sensing and nonlinear optics.   

In a traditional stationary WGM microresonator, the evanescent field extends only a short distance, on the order of the wavelength, beyond the resonator surface. \cite{arnold2009whispering} Plasmonic nanostructures enhance light–matter interaction by precisely positioning at the surface of a microresonator.\cite{ ahn2013demonstration, baaske2016optical, dantham2013label, shopova2011plasmonic, kang2017plasmon} Moreover, restoring the original WGM field distribution in such static systems poses a significant challenge, limiting their reusability and dynamic sensing capabilities.\cite{ahn2013demonstration, baaske2016optical, dantham2013label}

\begin{figure*}[h!]
  \centering
  \includegraphics[width=\linewidth]{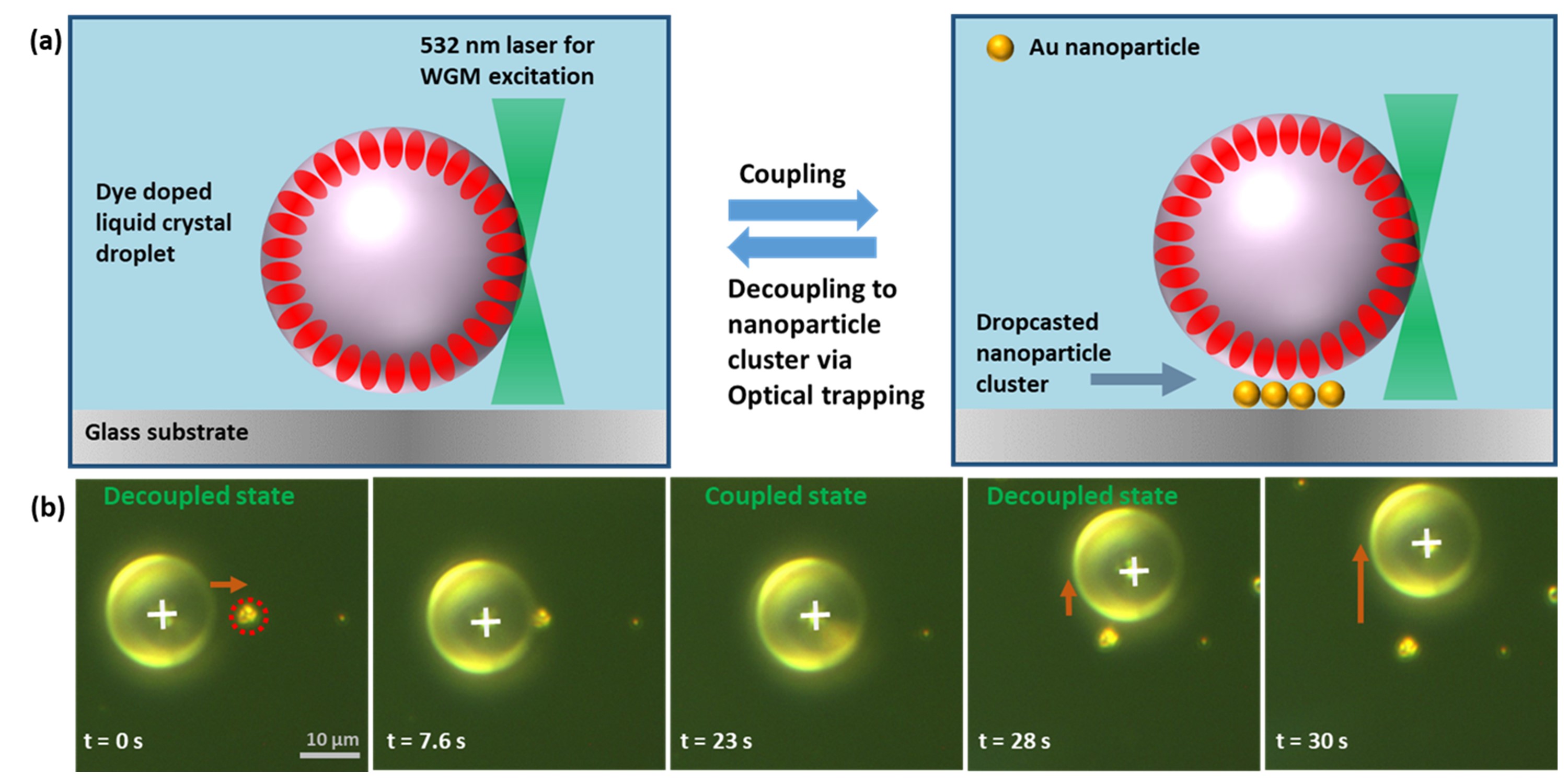}
  \caption{(a) Schematic illustration of whispering gallery mode (WGM) resonance in a dye-doped nematic microdroplet in its decoupled and coupled states relative to a plasmonic gold nanoparticle cluster drop-casted on a silica substrate. The droplet is excited by a 532 nm laser (green arrow) focused at the droplet–medium interface.  Red ellipsoids represent the electric field distribution of resonant WGMs. Bidirectional arrows represent the reversible coupling and decoupling process done by trapping with a 532 nm laser between the microdroplet and the gold nanoparticle cluster. The gold nanoparticle cluster, deposited on the substrate, is highlighted by the faded blue marking. (b) Time-sequence optical microscopy snapshots of a dye-doped nematic microdroplet under optical trapping, showing the controlled coupling–decoupling process. The cross symbol marks the focused laser beam position. At t = 0 s, the gold nanoparticle cluster (red circle) is positioned away from the droplet. As the trapped droplet is brought into proximity with the nanoparticle cluster (t = 7.6 s), it achieves parking over the cluster (t = 23 s). Subsequent nematic droplet moving away by trapping from the gold nanoparticle cluster leads to decoupling (t = 28 and 30 s).}
   \label{fig:figure01.jpg}
\end{figure*}

To address the above-mentioned challenges, we used an optical tweezers \cite{jones2015optical, grier2003revolution, ashkin1970acceleration, neuman2004optical, marago2013optical, shukla2023opto} system to trap and precisely position the nematic microdroplet-based WGM microresonator on a finite number of static gold nanoparticles dropcasted on the silica substrate. \cite{jonavs2017thermal, iftiquar2024observing,  shechter2020direct} This dynamic coupling approach enhances spatial flexibility and enables controlled interaction with localised plasmonic fields. \cite{kang2017plasmon, mao2023whispering} A schematic of the plasmon-coupled WGM of a dye-doped nematic microdroplet resonator is shown in Figure 1(a).  In our experiment, we use a 532 nm laser both  for optical trapping and for exciting the whispering-gallery modes (WGMs) of the dye-doped nematic microdroplet. For WGM excitation, the laser is focused at the rim of the droplet shown in Figure 1(a). However, in Figure 1(b), the laser is positioned at the center of the droplet to achieve stable trapping and parking. Thus, the apparent change in the laser position between the figures reflects its different roles: trapping at the center versus WGM excitation at the rim. Trapping the microdroplet and using a mechanical stage, the spatial position of the static gold nanoparticle cluster in the sample plane was adjusted to enable precise coupling with the microdroplet shown in Figure 1(b) (see supplementary video 1). We demonstrate that coupling dye-doped nematic microdroplets with a finite number of static gold nanoparticles leads to both whispering-gallery modes (WGMs) enhancement and red-shifting, indicating photonic–plasmonic interaction. The enhanced WGMs and the observed resonance shifts confirmed a localisation of WGM field. Using optical tweezers, the WGM emission spectrum of the nematic microdroplet remained reversibly tunable during the coupling and decoupling process. The optical trapping-based system allows the droplet to move and scan across the whole sample surface. We observed varying degrees of whispering-gallery mode (WGM) enhancement and spectral redshift depending on both the overall size of the gold nanoparticle clusters and the characteristic size of their constituent nanoparticles.  Previous studies have demonstrated limited plasmonic tuning in solid microresonators, where repeated coupling and decoupling produced resonance red shifts \cite{kang2017plasmon} In contrast, our dye-doped nematic liquid crystal microdroplet exhibits a tunable redshift during the coupling and decoupling process with a gold nanoparticle cluster, facilitated by an optical tweezer system. A quantitative comparison summarizing the magnitude of plasmonic tuning observed in solid versus liquid crystal microresonators is provided in Table 1, Supplementary Information S14. This cost-effective approach enables tunable redshifts and supports single-molecule detection, providing a versatile platform for advanced biochemical, environmental, and medical sensing, consistent with recent developments in hybrid photonic–plasmonic and biosensing technologies.\cite{wang2021liquid, dai2025dual, niu2024hollow, jing2023optimally, dai2024lossy}

\section{Experimental Section}
\subsection{Materials}
We have purchased 4-pentyl-4-biphenyl-carbonitrile (5CB) nematic liquid crystal, sodium dodecyl sulfate (SDS), and gold nanoparticles from Sigma-Aldrich. Sodium dodecyl sulfate (SDS) was used as an anionic surfactant that provides homeotropic anchoring. The wavelength of the laser light coincides with the absorption spectrum of the Nile blue chloride dye. Deionised water (resistance \( \geq 18.2\  M\Omega\cdot cm\)) obtained from the Milli-Q water purification system was used in all of the experiments. The gold nanoparticles have diameters of 100 nm, 200 nm and 400 nm.

\subsection{Sample Preparation}
A glass coverslip was cleaned with acetone, followed by drying. Next, 2\(\mu\)L of gold nanoparticle (AuNP) solutions of 100, 200, and 400 nm sizes each, prepared separately in different tubes and dispersed in ethanol, were drop-cast onto separate cleaned glass substrates. The evaporation of ethanol leads to the formation of well-separated gold nanoparticle clusters composed of varying numbers of nanoparticles of varying sizes due to drying-induced aggregation. After solvent evaporation, SEM images confirmed that while clusters of different sizes form, the individual nanoparticle size remains unchanged shown in Supplementry Information S13. These AuNPs remain firmly attached to the surface and do not diffuse back into the solution. We have prepared emulsions of dye-doped nematic liquid crystal in MQ water. For this taken about 2 \(\mu\)L of nematic liquid crystal 5CB + 0.11\% w/w Nile blue chloride dye + 0.085\% w/w (3 mM) sodium dodecyl sulfate was used and mixed using a vortexer. The mixing was performed using a Tarsons Spinix Vortex Shaker (Model No. 3020) operating at a speed of 3000 rpm for 60 seconds. These parameters were used to obtain droplets of a suitable size range for our experiments with a stable radial configuration. Since the liquid crystal is an organic material not dissolving in water, showing the stability of the emulsion and droplet size, it remains stable during the experiment \cite {jonavs2017thermal}. The dye is soluble in liquid crystal (LC) and has little effect on the LC director axis. 
During the droplet preparation it was observed that at and above 2 mM SDS concentration, droplets of various diameters exhibit a radial configuration, while lower concentrations result in partial or nonuniform configurations like monopolar and bipolar configurations. \cite{shechter2020direct, jonavs2017thermal}. These liquid crystal droplets formed at the lowest concentration, and we were below the critical micelle concentration (CMC) for SDS: 10 mM. A concentration of about 3 mM SDS was chosen to ensure consistent homeotropic anchoring. Taking a glass substrate on which gold nanoparticles were drop-casted, about 15 \(\mu\)L of mixture solution is enclosed in a sample chamber between two glass coverslips. This method produces distinct, well-separated microdroplets scattered throughout the fluid medium. The micro-droplets of a suitable size range are highly stationary after a few hours of sample preparation in a chamber having a thickness of 120 \(\mu\)m. The chamber is placed on the stage of our optical trapping microscope. The wavelength of the green laser falls in the excitation band of Nile blue, and the fluorescence emission from the sample is observed within a range of 570–700 nm. All observations were carried out at room temperature.

\subsection{Experimental Setup}
Microdroplets of suitable size were excited with focused laser light of wavelength 532 nm. A single-beam optical trap is generated by focusing a continuous laser beam (typically with power less than 5 mW) with an objective (magnification 50x, 0.5 NA), and the trapping beam is expanded with a beam expander to overfill the back aperture of the objective lens. A laser beam was focused near the droplet rim and acquired a fluorescence spectrum from the dyes in the nematic liquid crystal microdroplet, having an exposure time of 2 s. Fluorescent light emitted from the excited droplets was collected using the upright microscope objective that was used for focusing the trapping and subsequently analysed by an imaging spectrometer. The resonance spectra coming from the microdroplets were studied with the help of the Horiba iHR320 spectrometer. The minimum spectral resolution of the spectrometer is 0.26 nm for gratings of 300 lines/mm. A schematic of the experimental setup is shown in Supplementary Information S1. In the experiment, a laser power of 55 \(\mu\)W, measured at the sample plane without the objective lens, was used to excite the WGMs. Dark-field microscopy (DFM) is used to view metal nanoparticles. The condenser has a dark-field ring attached to it, which allows the white light to illuminate the sample obliquely. The majority of light passes straight through the sample. Only the scattered light from the sample will be captured by the objective if the objective’s numerical aperture is less than the dark field concentrator. To study the scattering spectra of plasmonic particles coupled with the liquid crystal microdroplets, DFM was connected to a spectrometer. A High-speed, low-profile motorised scanning stage (MLS203-1) having dimensions of 260 mm \(\times\)\ 230 mm \(\times\)\ 31 mm, giving a minimum achievable incremental movement of 100 nm, ordered from Thorlabs, was used for suitable parking.

\section{Results and discussion} 

\begin{figure*}[h!]
  \centering
  \includegraphics[width=\linewidth]{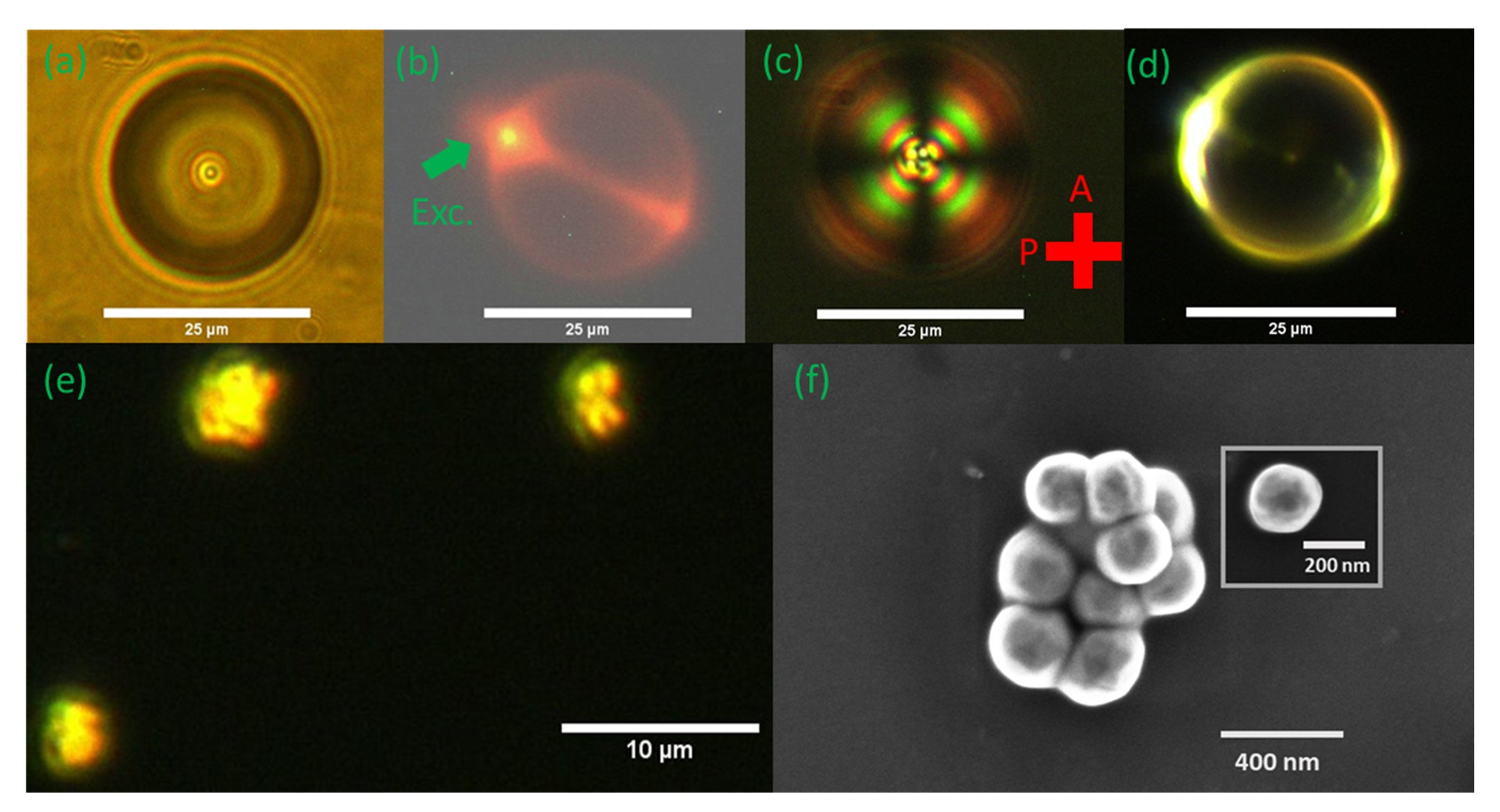}
  \caption{(a) Droplet as seen without polarisers. (b) A Nile-blue chloride dye-doped microdroplet illuminated by a focused 532 nm laser beam. The irradiation point is indicated by a green arrow. (c) Polarising optical microscope image of a 5CB nematic microdroplet in the presence of SDS solution at room temperature. A crossed polariser is shown on the bottom with red color. (d) Droplet visualised in dark field imaging. (e) Drop-casted nanoparticle cluster on silica substrate visualised in dark field. (f) Scanning electron microscopy (SEM) image of a gold nanoparticle cluster drop-casted on a silica substrate with a scale bar. The inset shows the size of an individual gold nanoparticle. All images of the same droplet are shown in figures (a), (b), (c), and (d).} 
  \label{fig:figure02.jpg}
\end{figure*} 

\subsection{Optical response of liquid crystal droplet and gold nanoparticle cluster morphology}
The bright field image of the liquid crystal microdroplet shown in Figure 2(a) indicates the sharp spherical boundary with refractive index contrast. We focused on an isolated microdroplet and excited it with green laser light just inside the SDS-liquid crystal interface, as the resonant modes are formed inside due to the sufficient contrast of refractive indices of the LC droplet and the host liquid. The arrow indicates the irradiation point in Fig. 2(b). The dye molecules emit fluorescence, and a thin, light ring around the circumference of the microdroplet is observed, indicating the excitation of WGM resonance. The restricted fluorescence within the microdroplet suggests that the dye molecules are not diffused in the surrounding medium. A polarised optical image of dye-doped 5CB liquid crystal and SDS droplets is obtained by a polariser and an analyser with an adjustable rotation angle shown in Figure 2(c). The different optical retardation from the centre to the perimeter causes colourful rings to form over the microdroplet. Each droplet’s center has four extinction branches aligned with the polariser axes. This implies that the long molecule axes are perpendicular to the liquid crystal-SDS interface, with a radial hedgehog defect in its centre, and that the director is radially symmetric. In the experiment, a dark field microscope was coupled with a spectrometer to measure the change in WGM spectra due to the coupling of gold nanoparticles with microdroplets. Figure 2(d) shows the dark field image of the same microdroplet. The different sizes of gold nanoparticle clusters formed that occurs as the ethanol suspension of gold nanoparticles evaporates from the glass substrate. The dark field image of the drop-casted gold nanoparticle cluster on the silica substrate is shown in Figure 2(e). The SEM images of the cluster of gold nanoparticles are shown in Figure 2(f), and the inset shows that the average size of the smallest unit of gold (Au) nanoparticle is 200 nm. The SEM image of a drop-casted gold nanoparticle cluster of different sizes is shown in supplementary information S2. 

\subsection{WGM mode characteristics}
WGMs are typically identified by two types of polarisations: transverse magnetic (TM) and transverse electric (TE), in which the electric field oscillates perpendicular to the sphere’s surface and parallel to it, respectively. The 5CB liquid crystal shows birefringence and has ordinary and extraordinary refractive indices that vary depending on the polarisation of incident light, i.e., either TE or TM mode. Since the droplet is in a radial director configuration, the optic axis points in the radial direction. The ordinary refractive index ( n$_o$) is experienced by TE-polarised modes, and the extraordinary refractive index ( n$_e$) is experienced by TM-polarised modes \cite{humar2009electrically}. It is also known that droplets with larger diameters support both TM and TE-polarised modes with higher mode values, while small LC microdroplets ( $<15 $ \(\mu\)m ) support only TM-polarised modes with the lowest radial mode number $ n = 1 $, which circulates closest to the cavity surface. Higher radial mode numbers always have a lower intensity and Q factor than lower mode numbers $ n = 1 $ \cite{humar2009electrically}. Also, the refractive index contrast between  n$_o$ and the surrounding medium is less in the TE-polarised mode, resulting in lower intensity and Q factor for a given radial mode number than the TM-polarised mode. In our experiments, the microdroplets of size (15 - 25)\(\mu\)m were used for WGM excitation and were precisely positioned over a finite number of gold nanoparticle clusters. The recorded WGM spectra display very small, low-intensity peaks corresponding to either higher-order TM or TE modes, consistent with earlier reports.\cite{humar2009electrically} Additional examples of such higher-order modes are provided in the Supplementary Information.

\subsection{Coupling of gold nanoparticle cluster with dye-doped microdroplet}

\begin{figure*}[h!]
  \centering
  \includegraphics[width=\linewidth]{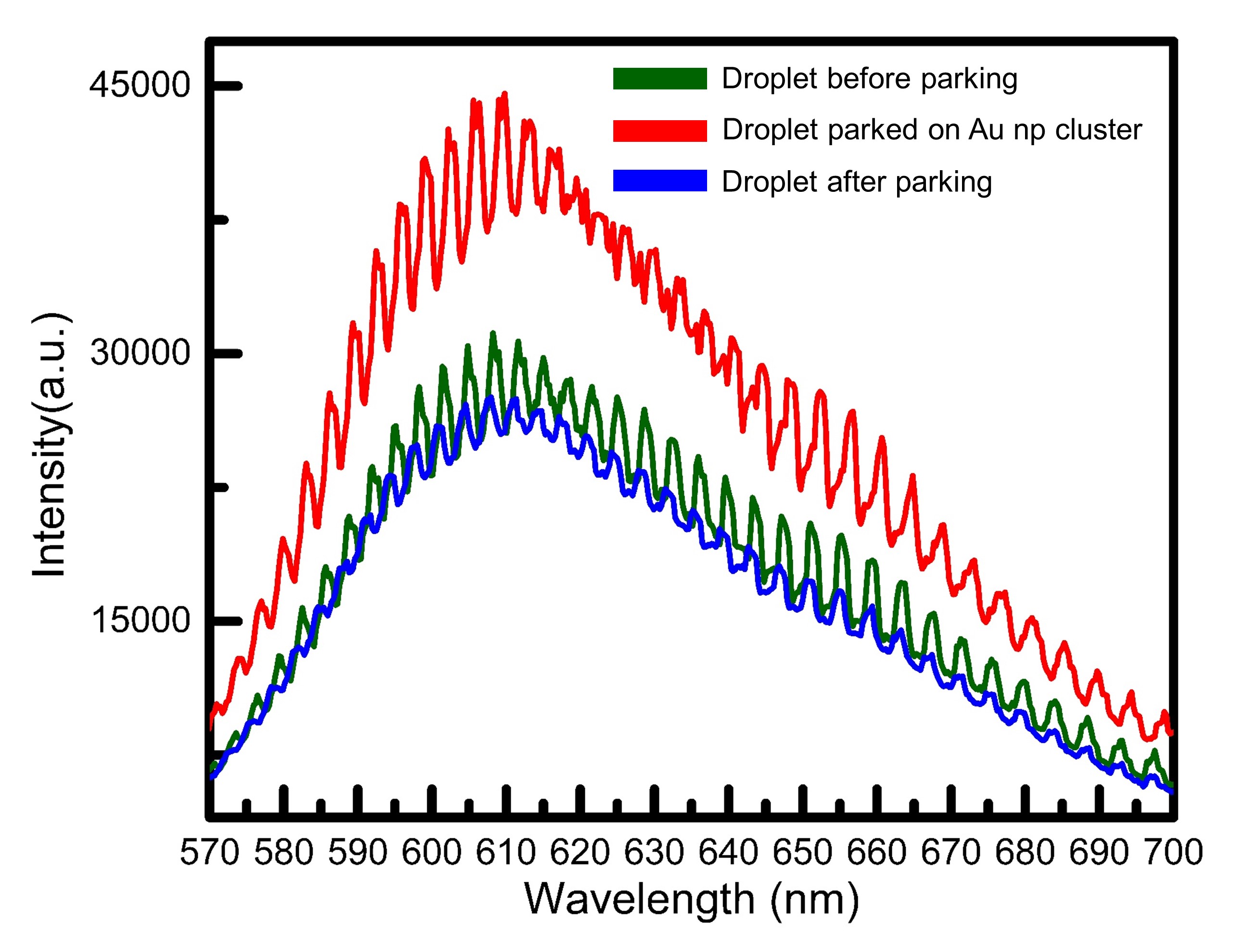}
  \caption{Whispering gallery mode (WGM) spectra of a dye-doped nematic microdroplet during coupling and decoupling interactions with a gold nanoparticle cluster. The green curve represents the WGM spectra of the isolated droplet before coupling. The red curve shows the WGM spectra when the droplet is coupled to the gold nanoparticle cluster, indicating enhanced fluorescence intensity due to plasmonic interaction. The blue curve corresponds to the decoupled state, where the droplet is once again isolated, showing the restoration of the almost original WGM spectrum.} 

  \label{fig:figure03.jpg}
\end{figure*}  

\subsubsection{Coupled vs. Uncoupled}
Figure 3 presents the intensity versus wavelength plot of the WGM resonance spectra for a dye-doped liquid crystal microdroplet (24 \(\mu\)m in diameter) before and during coupling with a gold nanoparticle cluster. The spectra, recorded in the wavelength range of 570–700 nm, show a marked difference in fluorescence background intensity and sustained WGM features. The uncoupled spectrum, shown in green, was collected by exciting the droplet before parking the droplet on the gold nanoparticle cluster. The coupled spectrum, shown in red, was recorded after the microdroplet was optically trapped at minimal power and parked onto the gold nanoparticle cluster, causing WGM emission with amplified fluorescence background. This confirms that the plasmonic near-field enhances the emission intensity without compromising the resonant feedback, validating the effectiveness of the optical coupling approach. Due to the precise positioning of the gold nanoparticle cluster within the evanescent optical field of the resonating whispering-gallery modes (WGMs) of the microdroplet, the local electric field near the metal surface is significantly enhanced via localised surface plasmon resonance (LSPR). This enhancement strengthens the electric field inside the microresonator, resulting in a highly amplified signal. The dye molecules, embedded within the liquid crystal microdroplet, reside well within the WGM’s evanescent field. When the dye molecule comes in the vicinity of a gold nanoparticle cluster, the interaction with LSPR modifies fluorescence behaviour and enhances the overall fluorescence intensity, and this power is efficiently trapped into the droplet, producing WGM-sustained spectra over an amplified fluorescence background. The fluorescence intensity increased by approximately 45 \% relative to its initial uncoupled value, as determined by integrating the emission spectrum over the wavelength range of 570–700 nm. The enhancement was observed in multiple coupling spectra measurements for the gold nanoparticle cluster. 

To further observe the effect of decoupling, the droplet was trapped and moved away from the gold nanoparticle cluster, and the WGM spectra of the isolated droplet were collected, shown by the blue colour in Figure 3. The fluorescence intensity of WGM modes in the case of after coupling has a slightly lower intensity compared to the WGM spectra before coupling. It is shown that the peak intensity of the modes decreases compared with the modes before coupling. However, due to continuous trapping and parking of droplets at suitable pump powers available for these experiments, degradation of the dye gain medium was observed. To confirm the decrease in fluorescence intensity after decoupling is due to dye photobleaching because of prolonged irradiation by optical tweezers, not because of the mechanical deformation of the droplet. We have performed a controlled experiment in which dye-doped LC droplets were trapped by optical tweezers for the same duration as the coupling–decoupling experiments without the gold nanoparticle cluster. WGM spectra were collected at different times, and obtained overlapping WGM modes, confirming no change in droplet size, while a gradual decrease in fluorescence intensity was observed, as shown in supplementry information S11. This indicates that the observed fluorescence decay arises mainly from the thermal effects of laser irradiation rather than mechanical disturbance from the tweezers. We further have verified photobleaching behavior of dye for different droplets shown in supplementary information S12. A 532 nm laser gives good trapping capability and overlaps with the absorption peak of Nile blue dye and thus causes photobleaching.\cite{jana2021impurity} 

\subsubsection{Cluster Size Effect}
We have found fluorescence enhancement for the smallest possible nanoparticle cluster found in the same sample, and we observed fluorescence intensity increased by approximately 24 per cent relative to before coupling, as shown in Supplementary Information S8. To further validate this, an experiment was conducted using a drop-casted gold nanoparticle cluster composed of nanoparticles with the smallest unit size having an average diameter of 100 nm. When a dye-doped microdroplet was optically parked on this cluster, a significantly amplified fluorescence background was observed, with sustained whispering-gallery modes (WGMs) visible in the emission spectrum. Upon integrating the emission over the 570–700 nm wavelength range, the fluorescence enhancement was found to be approximately 76 per cent compared to the initial intensity shown in Supplementary Information S3. Additionally, experiments were performed using gold nanoparticle clusters composed of larger nanoparticles (smallest unit average size 400 nm), which showed a moderate enhancement of roughly 37 per cent for parking a droplet of 18 \(\mu\)m diameter (Supplementary Information S9) on a gold nanoparticle cluster shown in Supplementary Information S10. The enhancement primarily depends on the plasmonic coupling strength, the nano gap between the nanoparticles in a cluster, and the scattering behavior. For 100 nm AuNPs, densely packed clusters with numerous nanogaps generate strong localized surface plasmon resonances (LSPRs), producing the highest enhancement (76\%). In a particular case where we performed the parking of a droplet on a smaller nanoparticle cluster having a basic unit of 200 nm containing only a few nanoparticles, leading to fewer interparticle junctions (“hot spots”), which weakens the near-field intensity and results in lower enhancement (24\%). On the other hand, for a relatively bigger cluster having more nano gaps (“hot spots”), we have observed an enhancement of 45\% in the Figure 3. However, the 400 nm AuNPs form larger aggregates with many coupled nanoparticles. These clusters support collective plasmonic modes and exhibit a larger scattering cross-section, promoting stronger coupling with the droplet’s whispering-gallery modes (WGMs). As a result, the 400 nm clusters yielded a higher value (37\%) compared to the 200 nm case. Other than this, as the basic unit size of the gold nanoparticle cluster increases, thermal effects induced during optical trapping of droplets and parking become significant \cite{shukla2025synchronized}. These effects can cause the gold nanoparticle cluster to adhere to the droplet surface, making it challenging to sustain WGMs and achieve consistent fluorescence enhancement, thereby requiring precise control over the trapping parameters. We also explored the use of different plasmonic geometries, particularly silver nanowires with an average diameter of roughly 350 nm prepared using the polyol method \cite{sun2003polyol}. When droplets were parked on these nanowires, whispering gallery modes sustained over highly amplified fluorescence signals were observed, with WGMs remaining well sustained over the enhanced background. Together, the microresonator, dye, and plasmonic nanoparticles produce stronger radiative enhancement than the plasmonic array or the resonator alone \cite{gartia2014injection, ahn2012photonic, ahn2013demonstration}. The morphology and size of the gold nanoparticle clusters with basic unit of 100/200/400 nm have been characterized quantitatively shown in supplementary information S13.

    \begin{figure*}[h!]
\centering
\includegraphics[width=\linewidth]{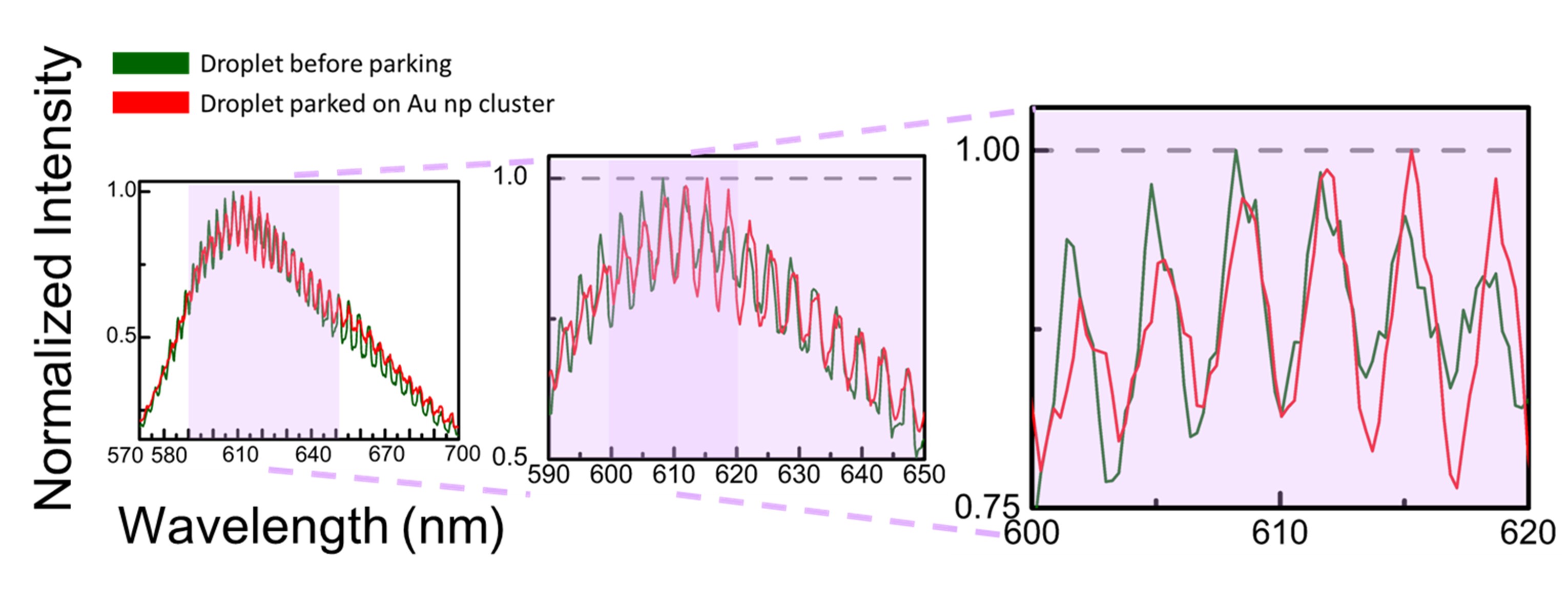}
\caption{A normalised intensity versus wavelength plot of whispering-gallery modes (WGMs) of a dye-doped nematic microdroplet before coupling (green) and during coupling (red) with a gold nanoparticle cluster. The left panel shows whispering-gallery modes over the broad fluorescence emission spectrum (570–700 nm) before and during coupling. The shaded light purple region is enlarged as indicated by dotted lines in the central panel (590–650 nm), highlighting mode spacing and the non-overlap of the highest intensity modes in decoupled and coupled states. The dark light-purple band in the central panel is further enlarged, as indicated by dotted lines, to show the regions of spectral magnification presented in the rightmost panel. This right panel zooms in on the 600–620 nm range, clearly showing the red shift of the highest intensity mode in the coupled state. A horizontal grey dotted line marks the peak normalised intensity to guide the eye in comparing the spectral shift of modes before and during coupling.} 
\label{fig:figure04.jpg}
\end{figure*}

\subsection{Whispering gallery modes of dye-doped microdroplet coupled to gold nanoparticle cluster}

\subsubsection{Drop-cast of gold nanoparticle} Prior studies have shown that nanostructures sized 200–300 nm provide strong field confinement and can be reliably fabricated, making them ideal for plasmonic coupling \cite{lin2015optimizing, kim2006nanowire, kim2009localized, ruesink2015perturbing}. This motivated us to drop-cast gold nanoparticles with an average size of 200 nm to achieve localised plasmonic fields. The drop casting method gives different sizes and shapes of the drop-casted gold nanoparticle cluster (dark field image shown in Figure 2e). The experiment was repeated by precisely parking the microdroplets on newly formed gold nanoparticle clusters, which were relatively larger in size compared to those used in Figure 3 of the previous section.

\subsubsection{Observation of Redshift}
The normalised WGM intensity plot of a 24 \(\mu\)m droplet for a wavelength range (570–700 nm) before and during coupling with a gold nanoparticle cluster, shown in Figure 4. To show the better visibility of resonance modes, we zoomed the wavelength range from 590 to 650 nm of the normalised intensity plot in the central panel indicates mode spacing and non-overlapping highest intensity modes. The central panel’s dark light purple band is highlighted with dotted lines and expanded in the rightmost panel, which zooms into the 600 to 620 nm spectral region. A grey dotted line indicates the highest value of normalised intensity drawn parallel to the wavelength axis. This magnified view clearly shows the redshift of the most intense mode in the coupled state. The highest normalised intensity mode before and during coupling touches the grey dotted line at different values of wavelength along the wavelength axis, 608.2 and 615.3 nm, respectively. We have repeated this parking experiment more than three times and observed a similar magnitude of shift. This indicates that during coupling, a red-shifted wavelength was observed. 

Compared to previously published plasmonic tuning values for solid microresonators, this value is two orders of magnitude higher \cite{kang2017plasmon}. More importantly, the spectral shift exceeds the free spectral range, meaning that the resonance wavelength can be shifted to any value. Comparing the tuning range with other techniques, Dhara et al. \cite{venkitesh2024two} recently have shown a blue shift due to a decrease in the effective refractive index value of liquid crystal droplets in an externally applied electric field. M. Humar et al. \cite{humar2009electrically} first showed that the tunability of liquid crystal droplets exceeds the free spectral range (FSR) in an external electric field. Kim et al. \cite{kang2017plasmon} report a 0.011 nm red shift of the TM modes of  WGM resonances at a 1310 nm excitation wavelength by repetitively coupling a 200 \(\mu\)m diameter solid microcavity to a 50 nm height with a 300 nm diameter single nanodisk on top of a 10 nm gold film and a 0.036 nm red shift for a double nanodisk. This indicated that upon increasing the number of nanodisks of similar size, the localisation of the field enhances, and a greater magnitude of shift is observed. We have included a table for a quantitative comparison summarising the magnitude of plasmonic tuning observed in solid versus liquid crystal microresonators, provided in supplementry information S14. 

\subsubsection{Red shift for different Cluster Size}
Apart from this, we have found the redshift in the case of the normalised intensity plot for the fluorescence enhancement data during coupling with a gold nanoparticle cluster, shown in Figure 3 in supplementary information S6. We also have observed that as the magnitude of the redshift increased, the fluorescence signal intensity dropped below the WGM fluorescence background measured before coupling. 
A more detailed mechanistic discussion is given, which clarifies the interplay between enhancement and quenching. The observed redshifts arise from perturbations of the droplet’s evanescent field by the gold nanoparticle cluster. The resulting fluorescence behavior is strongly dependent on the fluorophore–metal separation. When dye molecules are positioned within the enhanced electromagnetic field region (approximately 10–30 nm from the plasmonic surface), the radiative decay rate increases, leading to fluorescence enhancement. At sub-10 nm distances, however, non-radiative energy transfer to the metal becomes dominant and results in fluorescence quenching. This enhancement—quenching competition is well established. Novotny and co-workers experimentally demonstrated the continuous transition from fluorescence enhancement to quenching as the fluorophore-nanoparticle distance is varied. \cite{anger2006enhancement} A similar interplay between plasmonic field enhancement and additional loss channels has been highlighted in the review of biosensing with WGM. This review discusses how metallic nanostructures in the WGM evanescent field can simultaneously strengthen the local field and introduce non-radiative damping pathways, depending on their relative placement.\cite{toropov2021review} Furthermore, studies from our group on microsphere–waveguide coupling demonstrate how modifications in the local photonic environment—such as coupling to nearby dielectric waveguides—can modify WGM spectral characteristics through both field enhancement and perturbative losses.\cite{chikkaraddy2013microsphere} Their findings reinforce the idea that the balance between enhanced confinement and additional dissipative channels dictates the net spectral response.

Now we have again confirmed the red shift by parking the same microdroplet on different-sized drop-casted nanoparticle clusters shown in the supplementary information S7. We further verified the redshift by parking a liquid crystal droplet having a diameter of 22 \(\mu\)m on the gold nanoparticle clusters having a basic unit with an average size of 100 nm. The observed redshift corresponded to the free spectral range (FSR) of the droplet, as shown in Supplementary Information S4. By these observations we can say that the magnitude of the redshift changes for different sizes of gold nanoparticle clusters. In our experiments, different droplet sizes were selected for parking based on their availability in the sample that formed during the vortex of the emulsion. Slightly larger droplets were specifically chosen to facilitate easier parking over relatively smaller gold nanoparticle clusters. Repeated parking on different gold nanoparticle clusters gives a different magnitude of the redshift depending upon the strength of coupling. Since in our experiment we have no control over the size and shape of these drop-casted gold nanoparticle clusters, the magnitude of these red shifts varies for different clusters. The gold nanoparticle cluster localises the amount of light once it comes in the vicinity of the evanescent field of the resonating light along the circumference of the microdroplet, also known as evanescent field perturbation. This suggests that by aligning the gold nanoparticle cluster under study, a microdroplet can function as an efficient probe of the near-field that shows a different magnitude of shift depending upon coupling strength.

\begin{figure*}[h!]
  \centering
  \includegraphics[width=\linewidth]{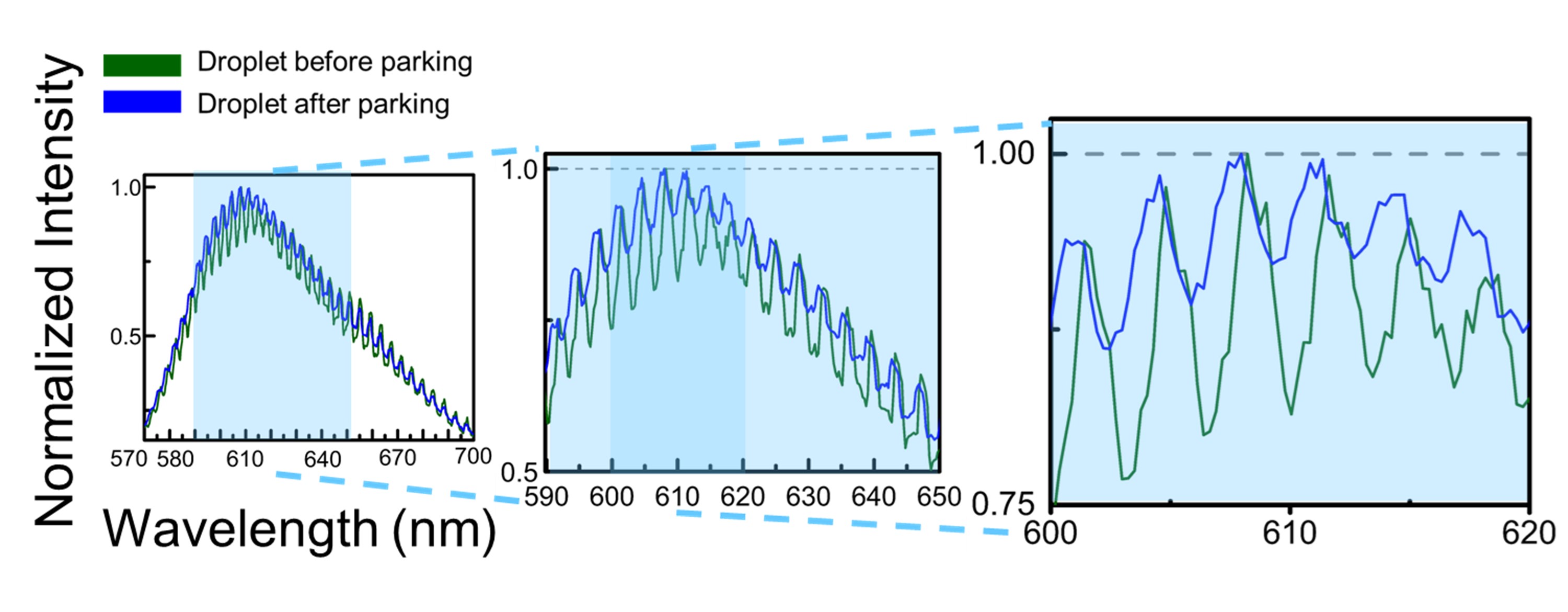}
  \caption{A normalised intensity versus wavelength plot of whispering-gallery modes (WGMs) of a dye-doped nematic microdroplet before (green) and after coupling (blue) with a gold nanoparticle cluster. The left panel shows whispering-gallery modes over the broad fluorescence emission spectrum (570–700 nm) before and after coupling. The shaded light sky-blue region is enlarged as indicated by dotted lines in the central panel (590–650 nm), highlighting mode spacing and the overlap of modes in decoupled states. The dark light sky-blue band in the central panel is further enlarged, as indicated by dotted lines, to show the regions of spectral magnification presented in the rightmost panel. This right panel zooms in on the 600–620 nm range, clearly showing the overlapping of individual modes before and after coupling. A horizontal grey dotted line marks the peak normalised intensity to guide the eye in comparing the reversibility of the highest intensity modes in decoupled states.}
  \label{fig:figure05.jpg}
\end{figure*}

\subsubsection{Q-Factor Analysis}
Using the resonance line width ($\Delta\lambda$) of the mode with the highest intensity having wavelength ($\lambda$), we calculated the Q factor ( \(Q = \frac{\lambda}{\Delta\lambda}\)) to quantitatively characterize the resonance spectra, where $\Delta\lambda$ is also known as the full width at half maxima of a mode. In Figure 4, the line width $\Delta\lambda$ corresponding to the resonance peak before coupling the microdroplet on the gold nanoparticle cluster (at $\lambda$ = 608.2 nm) is 1.87 nm, and the Q factor is roughly 325. When the droplet is precisely parked on the gold nanoparticle cluster, its line width $\Delta\lambda$ of the highest intensity mode (at $\lambda$ = 615.3 nm) is 1.98, and the Q factor is found to be roughly 310. The observed WGM linewidths indicate that the measured Q values are primarily limited by the intrinsic optical losses of the dye-doped nematic microdroplet. The Q-factor of the highest intensity mode during coupling slightly decreased as compared to before coupling of the microdroplet with the gold nanoparticle cluster. Plasmonic coupling introduces additional radiative and absorptive losses that have broadened the modes slightly compared to uncoupled droplets. This field enhancement, without significantly degrading the Q-factor, ensures localisation of the optical field. Future experiments with higher-resolution spectroscopy will allow a more precise determination of the intrinsic Q. \cite{humar2009electrically}
In addition to the main resonances, very small peaks with low intensity are also visible in the WGM mode that correspond to higher-order WGMs, arising from either TE or TM polarizations. \cite{humar2009electrically}

\subsection{Decoupling leads to reversibility of modes}
Figure 5, left panel is a normalised intensity plot of the WGM spectra of microdroplets before and after coupling with a gold nanoparticle cluster in 570-700 nm, shown in green and blue colours, respectively. To enhance visibility of the resonance modes, we enlarged the light sky-blue shaded region as a central panel (590–650 nm). This magnified view clearly illustrates the overlap of individual modes before and after coupling. A darker sky-blue band within this region is further expanded and shown in the rightmost panel, which focuses on the 600–620 nm spectral range. A horizontal grey dotted line marks the peak normalised intensity drawn parallel to the wavelength axis. The highest normalised intensity mode before and after coupling touches the grey dotted line and overlaps, indicating that the WGM modes show reversibility and are highly sensitive to changes in perturbation. The line width $\Delta\lambda$ corresponding to the resonance peak after coupling the microdroplet to the gold nanoparticle cluster remains the same as before coupling. The free spectral range (FSR), i.e., the separation of wavelengths between two consecutive modes denoted by $\delta\lambda$. Mathematically,
FSR = $( \delta\lambda = \frac{\lambda^2}{2\pi{n_r}R})$,
Where R is the radius of the microdroplet and $\lambda$ is the resonant wavelength, n$_r$ is the effective refractive index. Trapping may causes the slight increase in temperature, and the effective refractive index of the liquid crystal droplet decreases. The effective refractive index (n$_r$) is inversely proportional to FSR ($\delta\lambda$) and is expected to increase \cite{sofi2017electrical}. The distance between two consecutive modes is measured by the free spectral range (FSR); it is around 3.5 before and after coupling, indicating that the size of the droplet and the effective refractive index remain the same. As the trapping power was around 5 mW at the sample plane, which does not sufficiently cause any measurable change in the effective refractive index of the droplet\cite{jonavs2017thermal}. In our decoupling case, the observed shift is significantly smaller and is out of the limit of resolution of the spectrometer. 

The translational motion of the liquid crystal (LC) microdroplet toward the gold nanoparticle (AuNP) cluster is precisely controlled using optical tweezers. A focused 532 nm laser generates a gradient force that traps and moves the droplet via a computer-controlled motorised scanning stage (minimum step size: 100 nm). Minor local anchoring distortions may occur at the droplet–cluster contact region due to mechanical interaction, but no measurable macroscopic deformation is observed. The elastic restoring forces of the LC droplet cause the droplet to regain its spherical shape. If a temporary distortion occurs in the shape of the droplet, the WGM resonances would show clear shifts or mode splitting. But it is confirmed experimentally by the complete overlap of normalised WGM spectra recorded before and after trapping, indicating that the mechanical interaction does not alter the droplet geometry.
To further confirm the reversibility of the modes, we have performed highly controlled experiments by drop-casting gold nanoparticle clusters composed of nanoparticles with an average diameter of 100 nm. For the emission spectrum in the range of 570-700 nm, we have observed an equal number of overlapping modes before and after coupling, showing reversibility of WGM modes without any significant change in the FSR value, as shown in supplementary information S5. Thus, WGMs can be used to monitor the minute change in size and refractive index of dye-doped liquid crystal microdroplets in greater detail using spectroscopic measurements. 

\section{Conclusions}
In summary, we have demonstrated for the first time plasmon-coupled whispering gallery modes (WGMs) in dye-doped nematic liquid crystal microdroplets, with precise spatial control enabled by optical tweezers. The coupling between the WGM evanescent field of dye-doped liquid crystal microdroplets and gold nanoparticle clusters leads to significant fluorescence enhancement, indicating efficient near-field photonic-plasmonic interaction. The tunability of WGM resonances of dye-doped nematic microdroplets with plasmonic coupling is sufficiently larger in magnitude than previously known for solid microresonators \cite{kang2017plasmon}. Red shifts in the WGM spectrum, observed without significant degradation in Q-factor, highlight the system’s high sensitivity to nanoscale perturbations. Reversible coupling and decoupling, facilitated by optical tweezers, establish the tunability and reusability of this hybrid platform. Overlapping before and after coupling spectra confirm reversibility, which can be obtained by optimising trap parameters. Our results show that optimized plasmonic substrates—enabled by advanced fabrication technology—determine the baseline field enhancement, and fine control of interparticle spacing provides dynamic tunability of redshift, fluorescence enhancement, and mode splitting. This cost-effective approach enables tunable redshifts and supports single-molecule detection, providing a versatile platform for advanced sensing in biochemical, environmental, and medical applications.

\section*{Supporting Information}
Use \href{https://drive.google.com/drive/folders/173dq3Yh0UK0hg1AJExTuWyRt4aQ-MZH_?usp=sharing}{this link} to access the supporting information PDF and video related to the manuscript.

\section*{Acknowledgements}
The authors thank Dr Adarsh Vasista, Dr Sunny Tiwari, Dr Rahul Chand, Ashutosh Shukla, and Sneha Boby for the fruitful discussion. Sumant Pandey acknowledges the CSIR-UGC for providing a fellowship. This work was partially funded by AOARD (grant number FA2386-23-1-4054) G.V.P.K.

\bibliography{citations}

\end{document}